\documentclass[a4paper,fleqn,usenatbib]{mnras}

\usepackage[T1]{fontenc}
\usepackage{ae,aecompl}
\usepackage{graphicx}
\usepackage{amsmath}
\usepackage{amssymb}
\usepackage{color}

\title[Anthropic explanation of $\Lambda$ and galaxy
  formation]{Testing anthropic reasoning for the cosmological
  constant with a realistic galaxy formation model}
 
\author[Sudoh et al.]{
Takahiro Sudoh,$^{1}$\thanks{(sudoh@astron.s.u-tokyo.ac.jp)}
Tomonori Totani,$^{1}$
Ryu Makiya$^{2,3}$
and Masahiro Nagashima$^{4}$
\\
$^{1}$Department of Astronomy, School of Science, the University of Tokyo, Hongo, Tokyo 113-0033, Japan\\
$^{2}$Kavli Institute for the Physics and Mathematics of the Universe, Todai Institutes for \\Advanced Study, the University of Tokyo, Kashiwa, 277-8583 Japan (Kavli IPMU, WPI)\\
$^{3}$Max-Planck-Institut f\"{u}r Astrophysik, Karl-Schwarzschild Str. 1, D-85741 Garching, Germany\\
$^{4}$Faculty of Education, Bunkyo University, Koshigaya, Saitama 343-8511, Japan\\
}

\date{Accepted XXX. Received YYY; in original form ZZZ}

\pubyear{2016}

\begin{document}
\label{firstpage}
\pagerange{\pageref{firstpage}--\pageref{lastpage}}
\maketitle

\begin{abstract}
The anthropic principle is one of the possible explanations for the
cosmological constant ($\Lambda$) problem. In previous studies, a dark
halo mass threshold comparable with our Galaxy must be assumed in
galaxy formation to get a reasonably large probability of finding the
observed small value, $P(<$$\Lambda_{\rm obs})$, though stars are
found in much smaller galaxies as well.  Here we examine the anthropic
argument by using a semi-analytic model of cosmological galaxy formation,
which can reproduce many observations such as galaxy luminosity
functions.  We calculate the probability distribution of $\Lambda$ by
running the model code for a wide range of $\Lambda$, while other
cosmological parameters and model parameters for baryonic processes of
galaxy formation are kept constant.  Assuming that the prior
probability distribution is flat per unit $\Lambda$, and that the
number of observers is proportional to stellar mass, we find
$P(<$$\Lambda_{\rm obs}) = 6.7\%$ without introducing any galaxy mass
threshold. We also investigate the effect of metallicity; we find
$P(<$$\Lambda_{\rm obs}) = 9.0 \%$ if observers exist only in galaxies
whose metallicity is higher than the solar abundance.  If the number
of observers is proportional to metallicity, we find
$P(<$$\Lambda_{\rm obs}) = 9.7\%$. Since these probabilities are not
extremely small, we conclude that the anthropic argument is a viable
explanation, if the value of $\Lambda$ observed in our universe is
determined by a probability distribution.
\end{abstract}


\section{Introduction}
\label{sec:intro}
The early observational indications for non-vanishing cosmological
constant, $\Lambda$ \citep{Efstathiou1990,Fukugita1990,Yoshii1993,Krauss1995,Ostriker1995}, 
were further strengthened by type Ia supernova observations
\citep{Riess1998,Perlmutter1999}, and established by the WMAP data of
the cosmic microwave background radiation \citep{Spergel2003}, leading
to the standard $\Lambda$CDM model that are consistent with many
further high-precision observational tests until now \citep[see
  e.g. ][for reviews]{Frieman2008,Weinberg2013}.  The cosmological
constant is interpreted as the vacuum energy density, but
theoretically natural values expected by particle physics are larger
than the observed value $\Lambda_{\rm obs}$ by a factor of at least
$\sim 10^{55}$ \citep[see e.g. ][for theoretical
  reviews]{Weinberg1989,Carroll2001,Sahni2002,Caldwell2009}.  The
discrepancy becomes even much worse if we assume that the natural
value of the cosmological constant is set by the Planck energy
density, which is $10^{123}$ times larger than the energy density
corresponding to $\Lambda_{\rm obs}$. This is the so-called smallness
problem. Furthermore, $\Lambda$ is not exactly zero but has a finite
value, and we are living in a very special epoch when the energy
densities of matter and vacuum become comparable in the long history
of the universe.  This is the coincidence problem.  Many approaches
have been proposed, ranging from a new field called dark energy to
modification of the gravity theory, but there is no satisfactory
solution yet for this problem.

One approach to this problem is using the anthropic principle \citep[][see also
  \citealt{Banks1985}]{Barrow1986,Weinberg1987}.  If $\Lambda \ll - \Lambda_{\rm obs}$,
the universe would have collapsed much earlier than the present epoch
\citep{Barrow1986}. On the other hand,
if $\Lambda$ is positively too large, gravitational condensation of
matter is suppressed by an accelerated expansion of the background
universe, and hence no galaxy or intelligent life is
formed \citep{Weinberg1987}.  Such an idea is supported by some
theories about very early universe suggesting a possibility that there
is an ensemble of many multiverses and $\Lambda$ varies among
different multiverses. It is reasonable to expect a nearly flat prior
probability distribution $p(\Lambda)$ per unit $\Lambda$, because
$\Lambda$ of a habitable universe must be smaller than the
theoretically natural scale $\Lambda_{\rm th}$ by many orders of
magnitude.  If $p(\Lambda)$ is non-zero at $\Lambda = 0$, and
$dp/d\Lambda \sim \Lambda_{\rm th}^{-1}$, $p(\Lambda)$ would be
essentially constant within the range of $\Lambda$ for a habitable
universe. Then not only the smallness but also the coincidence problem
is solved because the probability of observing an absolute value
smaller than $|\Lambda|$ scales as $\propto |\Lambda|$.

The anthropic argument may not simply work if not only $\Lambda$ but
also other physical constants or quantities are varied in different
multiverses \citep{Tegmark1998,Aguirre2001,Graesser2004}, but in this
work we consider the simplest scenario that only $\Lambda$ changes in
a flat universe.  Recently, it has been shown that only $\Lambda$
changes with a flat $p(\Lambda)$ among different homogeneous patches
of the universe created by inflation, if the theory of gravity is
extended to allow any inhomogeneous initial conditions about spacetime
and matter, in contrast to general relativity in which the four
constraints in the Einstein field equations limit the possible initial
conditions \citep{Totani2016}.

If we know the expected number density of observers in a universe with
the cosmological constant $\Lambda$, $n(\Lambda)$, we can calculate
the probability distribution $P(\Lambda)$ per unit $\Lambda$ for an
observer realized in a universe, as
\begin{equation}
P(\Lambda)=\frac{n(\Lambda) p(\Lambda)}{\int_{0}^\infty n(\Lambda) 
  p(\Lambda)d\Lambda} \ ,
\label{eq1}
\end{equation}
where $n(\Lambda)$ is a comoving density scaled to an early epoch when
the effect of $\Lambda$ is negligible, to correct the difference of
late expansion factor by changing $\Lambda$.  Here we consider only
$\Lambda \ge 0$, because the extension of $n(\Lambda)$ to the negative
range is dependent on the formation time scale of an intelligent
observer, which is highly uncertain.  If it takes $\sim 5$ Gyr
(appearance of human being after the formation of the Earth),
$n(\Lambda)$ should rapidly drop below $\Lambda \lesssim -
\Lambda_{\rm obs}$, and hence ignoring the rather narrow range of $-\Lambda_{\rm obs} \lesssim \Lambda < 0$ does
not seriously affect the probability calculation \citep[][see also
  \citealt{Peacock2007}]{Weinberg1996}.  The probability to observe
$\Lambda$ smaller than the observed value is then
\begin{equation}
P(< \Lambda_{\rm obs}) \equiv \int_0^{\Lambda_{\rm obs}}
P(\Lambda) d\Lambda \ .
\end{equation}
If this is not small, the anthropic principle can be a viable
solution to the cosmological constant problem.

Previous studies indeed show that $P(<$$\Lambda_{\rm obs})$ is not
extremely small (typically 1--10\%), based on the structure formation
theory of the universe with cold dark matter (CDM)
\citep{Efstathiou1995,Martel1998,Garriga2000,Peacock2007}.  In these
studies the total dark matter mass included in collapsed dark haloes
was used as an estimator of $n(\Lambda)$, and it was calculated
analytically by formulations like the Press-Schechter theory, but
baryonic physics related to formation of galaxies (e.g., gas
cooling, star formation, supernova feedback, starbursts by galaxy
mergers, metal production and chemical evolution) were not considered,
though it should also be important to estimate $n(\Lambda)$.
Furthermore, these studies assumed a minimum mass threshold for dark
haloes that can harbor life, with a value similar to that of our
Galaxy, to get a probability $P(<$$\Lambda_{\rm obs})$ that is not
extremely small.  However, such a treatment is clearly ad hoc, and we
know that the mass distribution of galaxies extends to dwarf galaxies
that are smaller than our Galaxy by a factor of 1000.

In this work, we calculate $n(\Lambda)$ in a wide range of $\Lambda$
by using a semi-analytic model of galaxy formation in the framework
of cosmological structure formation in a CDM universe.
In such a model, the baryonic processes of galaxy formation mentioned above are
taken into account, and model parameters are determined to reproduce a
variety of observed data including luminosity functions, galaxy number
counts, and several empirical relationships like the Tully-Fisher
relation \citep[see e.g.][for a review]{Baugh2006}. The aim of this work
is to examine whether a reasonably large $P(<$$\Lambda_{\rm obs})$ is
obtained without an ad hoc galaxy mass threshold, when we take into
account realistic physical processes of galaxy formation.  It is known
that the faint-end slope of galaxy luminosity function is flatter than
that of dark haloes, and supernova feedback is believed as the primary
mechanism working preferentially in low mass haloes to reduce the
amount of stars.  This would have a similar effect to the mass
threshold in the previous studies, but a quantitative computation is
necessary, which is the distinctive feature of this work.

We first calculate $n(\Lambda)$ assuming that the number of life
systems in a galaxy, $N_{\rm life}$, is proportional to stellar mass
of the galaxy, $M_*$. However, formation of a star is not a sufficient
condition of a habitable system for an observer. Clearly, we do not
expect formation of a terrestrial planet or life around a zero
metallicity star. Then there should be a metallicity dependence for
the probability of finding an observer in a stellar system. It is
known that massive galaxies generally have high metallicities, and
this trend is reproduced by galaxy formation models. Then
high-metallicity preference of life formation may also work as an
effective threshold in galaxy mass, and hence help the anthropic
argument of $\Lambda$. Metallicity evolution of galaxies is calculated
in the galaxy formation model used in this work, and we will also
study this effect quantitatively.

The outline of this paper is as follows. In Section \ref{sec:methods},
we describe the galaxy formation model that we use, and methods of
calculating $n(\Lambda)$ in universes with different
$\Lambda$. Section \ref{sec:results} presents our main results,
followed by a summary in Section \ref{sec:conclusion}.  We adopt the
"Planck+WP'' values reported in Table 2 of \citet{Planck} for the
cosmological parameters of the present universe that we observe.

\section{Methods}
\label{sec:methods}

\subsection{Semi-analytic modeling of galaxy formation}

In this work we use a semi-analytic galaxy formation model of
\citet[][hereafter NY04]{Mitaka}, called the Mitaka model.  A
semi-analytic modeling of cosmological galaxy formation starts from
producing a mock catalog of dark matter haloes and their merger
histories in the past.  A mock sample is constructed by producing many
dark haloes with various masses at an output redshift.  The merger
history of a dark halo (so-called merger tree) is generated by a
Monte-Carlo calculation based on the extended Press-Schechter (PS)
theory. Then they are summed up with an weight so that their number
density obeys to the dark halo mass function. In NY04, the halo mass
function was calculated by a fitting formula to $N-$body simulation
results, which is different from the analytic PS mass function by a
factor of at most 1.7. It is highly uncertain whether this fitting
formula holds for a value of $\Lambda$ quite different from that of
the standard $\Lambda$CDM model, and such a high precision correction
is not crucial in this work. Therefore we simply use the PS mass
function in this work. We
produce about 100 Monte Carlo realisations of dark halo merger trees
within a dark matter mass interval of 0.1 dex at an output redshift.

It should be noted that, instead of using the Monte-Carlo approach,
another way of generating a mock halo catalog is to directly sample
haloes and their merger histories from a cosmological $N-$body simulation
\citep[e.g.,][]{Makiya2016}. Though this gives a more accurate halo
mass function and merger histories, it is necessary to run many $N-$body
simulations for various values of $\Lambda$ for our purpose,
which would be an unrealistically high computational cost.  

Baryonic processes such as star formation, supernova feedback, and
galaxy mergers are then calculated in these haloes.  There are
phenomenological model parameters for various baryonic processes, and
they are determined to reproduce the observed luminosity functions and
cold gas mass fraction of galaxies at $z=0$.  Predictions of this
model are in agreement with not only various properties of $z=0$
galaxies, but also those at high redshifts, such as cosmic star
formation history and luminosity functions of Lyman break galaxies and
Lyman $\alpha$ emitters
\citep{Kashikawa2006,Kobayashi2007,Kobayashi2010}.  It should be noted
that these model parameters are for physical processes inside
collapsed dark haloes, and they are dependent only on physical
properties of a halo (e.g., mass and size), without direct dependence
on cosmological parameters. Therefore we fix the baryonic model
parameters when we try various values of $\Lambda$.

\subsection{Estimating the Amount of Observers, $n(\Lambda)$}
\label{subsec:N}

The cosmological parameters affect the results of the Mitaka model by
the dark halo mass function and their merger histories.  It should be
noted that the standard cosmological parameters, such as $H_0$,
$\Omega_M$, $\Omega_\Lambda$, and $\sigma_8$ which are inputs to the
Mitaka model, depend on the definition of the present time, i.e.,
$z=0$. In this work the flat geometry is always assumed
  after inflation in the early universe, according to the standard
  paradigm. As argued in Sec. \ref{sec:intro}, we assume that
only $\Lambda$ is changed in different multiverses, and hence all
other physical quantities do not change in the early universe when
$\Lambda$ is negligible. Here we define the initial epoch by the
cosmic time slightly after the recombination, $t_i = 4.7 \times 10^5$
yr, corresponding to $z= 1000$ in our universe. We calculate physical
densities of photons ($\rho_\gamma$), baryons ($\rho_B$), and dark
matter ($\rho_M$), and the amplitude of the matter density fluctuation
($\sigma_8$) at $t = t_i$ by solving the Friedmann equation and the
linear perturbation theory with the cosmological parameters of our
universe. Then these quantities, i.e.,
  $\rho_\gamma(t_i),\rho_B(t_i),\rho_M(t_i),$ and $\sigma_8(t_i)$, are
  set to be the same among different multiverses.

We need to consider about an output age of the universe that is
appropriate to calculate $n(\Lambda)$.  The age of the present
universe (13.8 Gyr), the age of the Sun (4.6 Gyr), and the
main-sequence life time of the Sun (10 Gyr) are all similar.  A
typical time scale for a life to evolve into an intelligent observer
may also be similar, though we know only one example.  Ideally, we
should integrate over time all stars that can harbor an intelligent
life, and here we estimate this by the total stellar mass density at
an age of the universe when the majority of cooled gas is already
converted into stars and hence significant more star formation is not
expected in future. In the present universe, the age 13.8 Gyr roughly
satisfies this condition, because the peak of cosmic star formation
($z \sim 1$--3) has already passed.  If we consider a $\Lambda$ value
larger than the observed value, gravitational collapses of dark haloes
occur only in earlier epochs, and hence we can use the same age. In
the limit of $\Lambda \rightarrow 0$, though larger mass dark haloes
would be formed at later epochs, star formation is not expected in
haloes that are much more massive than our Galaxy.  This is because the
gas cooling time becomes much longer than the age of the universe, due
to high temperature and low density, as seen in warm-hot intergalactic
medium and intracluster medium in the present universe. Therefore we
calculate the number of observers $n(\Lambda)$ at a fixed age of $t_0
= 15$ Gyr from the Big Bang for any value of $\Lambda$, and choice of
a different output age will be tested later. This defines the zero
point of redshift $z$ for a given value of $\Lambda$, and we can
calculate cosmological parameters of $H_0$, $\Omega_M$,
$\Omega_\Lambda$, and $\sigma_8$ at $t = t_0$ by evolving the universe
from $t = t_i$ to $t_0$.  These are used as inputs to the Mitaka model
calculation.

\begin{figure}
\includegraphics[height=1\columnwidth,width=0.9\columnwidth,angle=270]{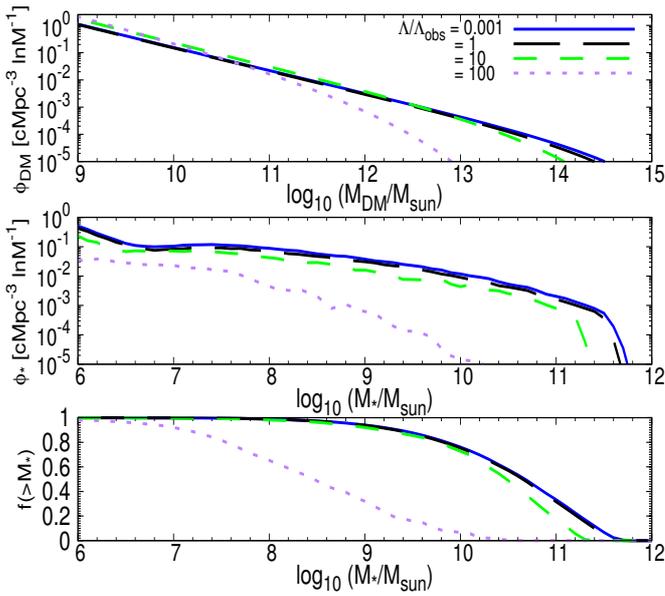}
    \caption{Dark matter mass function of dark haloes
        [$\phi_{\rm DM}(M_{\rm DM})$, top] and stellar mass function
        of galaxies [$\phi_{\rm *}(M_*)$, middle] are shown, for
        representative values of $\Lambda$ at $t_0 = 15$ Gyr. Here,
        mass functions (i.e., number density) are defined in
        logarithmic mass intervals, and hence are shown in units of
        per comoving Mpc$^3$ (cMpc$^3$) and $\ln M$.  In the bottom
        panel, the fraction of cosmic stellar mass contained in
        galaxies whose stellar masses are larger than $M_*$ is shown.}
    \label{figMF}
\end{figure}

The shape of power spectrum $P(k)$ of linear matter density
fluctuation as a function of the comoving wavenumber $k$ is necessary
as another input to the Mitaka model.  Here we consistently use the
$P(k)$ shape of our universe at $z=0$ calculated by the code CAMB
\citep{CAMB} for any value of $\Lambda$, because the shape of $P(k)$
in the linear regime does not evolve significantly after baryon
density fluctuation catches up that of dark matter, which occurs
shortly after the recombination. To calculate dark halo mass function,
their virial radii, and merger histories by the PS and extended PS
theories, we need to compute key quantities of the spherical collapse
model: $\delta_{\rm c}$, the fractional overdensity extrapolated
linearly to a time when a spherically symmetric fluctuation collapses
to a dark halo, and $\theta_{\rm v}$, the non-linear overdensity ($= 1
+ \delta$) of a collapsed and virialized object. We calculated these
for various values of $\Lambda$, by analytic
formulations given in \citet{Nakamura1997} for the case of a flat
universe. Calculated dark halo mass functions are shown
  in the top panel of Figure \ref{figMF} for some values of $\Lambda$.

From the outputs by the Mitaka model, we can compute cosmic densities of
various quantities (e.g., galaxy number or stellar mass) at the fixed
output age of $t_0$.  However, if we change $\Lambda$, it also changes
the expansion factor $a_0/a_i$ from $t = t_i$ to $t_0$.  For
calculation of $n(\Lambda)$ we need to consider an amount in a fixed
physical volume defined at the initial epoch, which is not affected by
$\Lambda$.  Therefore, we calculate $n(\Lambda)$ from the comoving
densities scaled to the initial epoch of $t = t_i$, which is denoted
with a subscript ``$i$'', e.g., $\rho_{*,i} \equiv \rho_*(t_0) (a_0/a_i)^3$
for stellar mass density.

Then we can calculate $P(\Lambda)$ from $n(\Lambda)$ and the prior
distribution $p(\Lambda)$.  We assume a flat distribution for
$p(\Lambda)$ only in $\Lambda \ge 0$, which is reasonable as discussed
in \S \ref{sec:intro} and assumed in previous studies
\citep[e.g.,][]{Martel1998}.

\section{Results}
\label{sec:results}

\subsection{Effects of Baryonic Processes}

\begin{figure}
	\includegraphics[angle=270,width=\columnwidth]{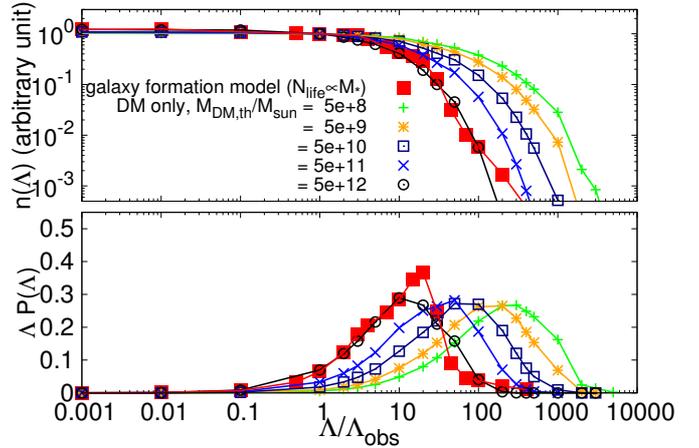}
    \caption{The number of observers $n(\Lambda)$ (top) and the
      probability distribution per unit $\ln \Lambda$, $\Lambda
      P(\Lambda)$ (bottom), are shown. The red filled squares are for
      the calculation using the Mitaka galaxy formation model,
      assuming that the number of observers is proportional to stellar
      mass ($N_{\rm life} \propto M_*$).  For comparison,
      calculations using only dark halo mass are also shown for
      several values of dark matter mass threshold, $M_{\rm DM, th}$.
      The probability distribution is normalized by $\int_0^\infty
      P(\Lambda) d\Lambda = 1$, while $n(\Lambda)$ is normalized to
      unity at $\Lambda/\Lambda_{\rm obs} = 1$.
}
    \label{figPS}
\end{figure}

First we calculate $n(\Lambda)$ only using dark halo masses, without
baryonic physics such as star formation, for comparison with previous
studies.  Figure \ref{figPS} shows the expected number of observers
$n(\Lambda)$ (top panel) and the probability distribution per unit
$\ln \Lambda$, i.e., $\Lambda P(\Lambda)$ (bottom panel), assuming
that $n(\Lambda)$ is proportional to the total dark matter mass of
collapsed haloes calculated by the PS mass function at $t_0$ = 15 Gyr.
Here, we introduced a mass threshold $M_{\rm DM, th}$ for dark haloes
and only haloes more massive than this are taken into account.  Results
for several values of $M_{\rm DM, th}$ are shown in the figure, and
the median of $P(\Lambda)$ and the probability $P(<$$\Lambda_{\rm
  obs})$ are summerized in Table \ref{tablePS}. If we set the
threshold close to our Galaxy, $M_{\rm DM, th} = 5\times 10^{11}\ {\rm
  M_\odot}$ \citep{Xue2008}, we find the probability
$P(<$$\Lambda_{\rm obs}) = 3.4$\% that is not extremely small, in
agreement with \citet{Martel1998}.  However, if we take smaller
thresholds, the probability $P(<$$\Lambda_{\rm obs})$ rapidly
decreases to 0.58\% for $M_{\rm DM, th} = 5\times 10^8\ {\rm
  M_\odot}$.

The main result of this paper, i.e., the probability distribution
calculated by the galaxy formation model assuming that the number of
life systems $N_{\rm life}$ is proportional to stellar mass $M_*$ in a
galaxy (and hence $n(\Lambda)$ proportional to the comoving stellar
mass density, $\rho_{*,i}$), but without any threshold about dark halo
mass, is also shown in Fig.  \ref{figPS}.  Compared to the results by
the PS mass function only, we find that $n(\Lambda)$ drops faster with
increasing $\Lambda$ even than the case of the largest $M_{\rm DM, th}
= 5 \times 10^{12} M_\odot$. When $\Lambda$ is large, only small mass
haloes can collapse at early epochs. Therefore this result indicates
that there must be a mechanism in the galaxy formation model to
suppress star formation in low mass dark haloes. Indeed, the supernova
feedback is commonly incorporated in galaxy formation models, and it
suppresses star formation in low mass haloes where interstellar gas is
easily expelled by the energy/momentum input from supernovae. This is
essential to make the faint-end of galaxy luminosity functions flatter
than that of PS mass function and to match the observed
data. This can be seen quantitatively in
  Fig. \ref{figMF}, where we present dark halo mass functions and
  galaxy stellar mass functions for some values of $\Lambda$. We find the median of $\Lambda/\Lambda_{\rm obs}$ in $P(\Lambda)$ to be 11 and the
probability $P(<$$\Lambda_{\rm obs}) = $ 6.7\%, which is not very
small without an ad hoc dark halo mass threshold.

\begin{figure}
	\includegraphics[angle=270,width=\columnwidth]{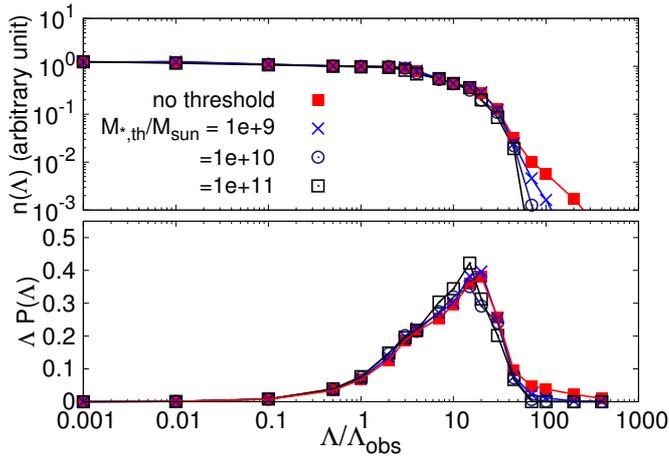}
    \caption{The same as Fig. \ref{figPS}, but for the results using
      the galaxy formation model ($N_{\rm life} \propto M_*$)
      with various values of stellar mass threshold $M_{\rm *, th}$.
}
    \label{figM}
\end{figure}

To further examine the effect of mass threshold in the galaxy
formation model, we show $n(\Lambda)$ and $P(\Lambda)$ using the
galaxy formation model but assuming that only galaxies more massive
than a stellar mass threshold $M_{\rm *, th}$ can harbor an observer,
in Figure \ref{figM}. This plot shows that changing $M_{\rm *, th}$
has little effects. The probability $P(<$$\Lambda_{\rm obs})$
increases only slightly to 8.0\% for $M_{\rm *, th} =
10^{11}{\rm\ M_\odot}$, compared with the no threshold case. This
confirms that the amount of stars in low mass galaxies is not
significant in the galaxy formation model thanks to the supernova
feedback. To show this quantitatively, the fraction
  $f(>$$M_*)$ of cosmic stellar mass contained in galaxies whose
  stellar masses exceed $M_*$ is shown in the bottom panel of Figure
  \ref{figMF}. When $\Lambda/\Lambda_{\rm obs} = 100$, only small
  galaxies of $M_* \lesssim 10^9 M_\odot$ can be formed, but the
  stellar mass fraction contained in such galaxies is small in the
  $\Lambda = \Lambda_{\rm obs}$ universe because of the feedback.

\begin{table}
	\centering
	\caption{The median of the probability distribution
          $P(\Lambda)$ and the probability of finding $\Lambda$
          smaller than the observed value, $P(<$$\Lambda_{\rm obs})$,
          as a function of the mass threshold of dark haloes, $M_{\rm
            DM, th}$. These results assume that the number of
          observers is proportional to the amount of dark matter mass
          in collapsed haloes, without taking into account galaxy
          formation.}
	\label{tablePS}
	\begin{tabular}{cccccc} 
		\hline
	        $M_{\rm DM, th}$ [$\rm M_\odot$] 
   & $5 \cdot 10^8$ & $5 \cdot 10^9$ & $5 \cdot 10^{10}$ 
   & $5 \cdot 10^{11}$ & $5 \cdot 10^{12}$\\
   median $\Lambda/\Lambda_{\rm obs}$ & 197 & 112 & 55.6 & 27.3 & 10.9 \\
		$P(<$$\Lambda_{\rm obs})$ [\%] & 0.58 & 0.97 & 1.9 & 3.4 & 7.7\\
		\hline
	\end{tabular}
\end{table}

\subsection{Dependence on Metallicity and Output Age}

It is reasonable to expect that metallicity affects the formation of
observers, because earth-like planets are composed of heavy elements
and known life systems on the earth need various heavy elements.  A
strength of using the galaxy formation model is that this model
provides metallicity of individual galaxies, and this enables us to
investigate the effects of metallicity. We test this using two simple
modelling: one is introducing a metallicity threshold for formation of
a habitable planet, and the other is assuming that the number of life
systems in a galaxy is proportional to $N_{\rm life} \propto M_* Z$
(and hence $n(\Lambda)$ proportional to the comoving density of $M_*
Z$, $\rho_{*Z,i}$), where $Z$ is metallicity (mass fraction of heavy
elements in all baryonic matter).

\begin{figure}
	\includegraphics[angle=270,width=\columnwidth]{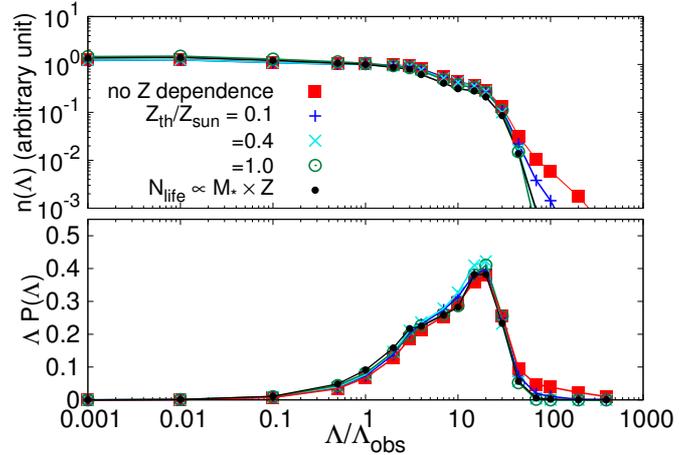}
    \caption{The same as Fig. \ref{figM}, but showing dependence on
      metallicity. Models assuming a metallicity threshold $Z_{\rm
        th}$ for an observer to exist are shown for some values of
      $Z_{\rm th}$.  Another model assuming that the number of
      observers $N_{\rm life}$ is proportional to $M_* Z$ in a galaxy
      is also shown.
}
    \label{figZ}
\end{figure}

Figure \ref{figZ} and Table \ref{tableZ} show the results of
introducing a metallicity thresholds $Z_{\rm th}$, under which no life
is assumed to exist. The dependence on $Z_{\rm th}$ is not strong;
$P(<$$\Lambda_{\rm obs})$ is slightly increased from 6.7\% to 9.0 $\%$
by increasing $Z_{\rm th}$ from zero to $Z_\odot$.  This implies that
the stellar mass in low metallicity galaxies is not a significant
fraction.  This is indeed expected from the well-known
mass-metallicity relation
\citep{Garnett2002,Tremonti2004,Savaglio2005,Erb2006,Lee2006};
metallicity is tightly correlated with stellar mass of galaxies, and
the stellar mass in low mass (i.e., low metallicity) galaxies is not a
significant fraction in the universe (see the previous section).  The
result for the case of $N_{\rm life} \propto M_* Z$ is also shown in
Figure $\ref{figZ}$. We find $P(<$$\Lambda_{\rm obs})=9.7\%$, which is
the largest among all the calculations tried in this work, though the
increase from the case of no metallicity effect (6.7\%) is modest.

\begin{table}
	\centering
	\caption{The same as Table \ref{tablePS}, but showing the
          dependence on the metallicity threshold $Z_{\rm th}$ in the
          calculations using the galaxy formation model, assuming
          $N_{\rm life} \propto M_*$ in galaxies whose
          metallicity is larger than $Z_{\rm th}$. }
	\label{tableZ}
	\begin{tabular}{cccccccc} 
		\hline
		$Z_{\rm th}$ [$Z_\odot$] & 0 &0.1 & 0.2 & 0.4 & 0.6 & 0.8 & 1.0\\
  median $\Lambda/\Lambda_{\rm obs}$ & 11 & 9.7 & 9.4 & 9.3 & 9.3 & 9.2 & 9.1\\
		 $P(<$$\Lambda_{\rm obs})$ [\%] & 6.7 &7.3 & 7.5 & 7.7 & 8.0 & 8.4 & 9.0 \\
		\hline
	\end{tabular}
\end{table}
		
\begin{table}
	\centering
	\caption{The same as Table \ref{tablePS}, but showing the
          dependence on the output age $t_0$, for the case of using
          the galaxy formation model assuming that the number of
          observers $N_{\rm life}$ is proportional to either $M_*$ or
          $M_* Z$ in a galaxy. }
	\label{tableage}
	\begin{tabular}{cccc}
		\hline
		$t_0$ [Gyr] & 10 & 15 & 20 \\
                \hline
                \multicolumn{4}{c}{Models assuming 
                    $N_{\rm life} \propto M_*$} \\
                \hline
  		median $\Lambda/\Lambda_{\rm obs}$ & 12 & 11 & 12\\
		$P(<$$\Lambda_{\rm obs})$ [\%] & 6.3 & 6.7 & 6.2\\
                \hline
                \multicolumn{4}{c}{Models assuming 
                    $N_{\rm life} \propto M_* Z$} \\
                \hline
   		median $\Lambda/\Lambda_{\rm obs}$ & 9.3 & 8.4 & 9.1\\
		$P(<$$\Lambda_{\rm obs})$ [\%] & 8.5 & 9.7 & 9.2\\
		\hline
	\end{tabular}
\end{table}

Finally, we check the dependence of our results on the output age of
the galaxy formation model, $t_0$, for which 15 Gyr was adopted in our
standard calculation.  In Table \ref{tableage} we show the results for
$t_0 = 10$, 15, 20 Gyr, for the models assuming $N_{\rm life} \propto
M_*$ or $M_* Z$. The dependence on $t_0$ is indeed small, well within
theoretical uncertainties.  This confirms that our results are not
significantly changed by a choice of $t_0$ in a reasonable range.

\section{Summary}
\label{sec:conclusion}

In this work we examined the anthropic argument to explain the
cosmological constant problem by a semi-analytic model of
cosmological galaxy formation, assuming that an observable universe is
created with a variable value of $\Lambda$ obeying to a nearly flat prior
probability distribution per unit $\Lambda$, while any other physical
parameter does not change.  The galaxy formation model used here
produces a mock catalog of dark matter haloes and their merger history
by Monte-Carlo simulations based on the structure formation theory in
the $\Lambda$CDM universe. Various astrophysical processes such as gas
cooling, star formation, and supernova feedback are phenomenologicaly
modeled to produce galaxies in dark haloes with physical quantities
such as stellar mass and metallicity. Astrophysical model parameters
have been determined to reproduce various observed data, and they are
assumed not to change for different $\Lambda$.

Assuming that the number of observers $N_{\rm life}$ is proportional
to stellar mass $M_*$ in a galaxy, we find a median in the probability
distribution $P(\Lambda)$ to be $\Lambda/\Lambda_{\rm obs} = 11.0$,
and the probability of finding $\Lambda \le \Lambda_{\rm obs}$ to be
$P(<\Lambda_{\rm obs})$ = 6.7\%. It should be noted that we obtained
this result without introducing any galaxy mass threshold, which is in
contrast to previous results based only on the formation history of
dark matter haloes.  Using the PS formalism and assuming that the
number of observers is proportional to dark matter mass of collapsed
haloes, we confirmed the previous results that a mass threshold close
to our Galaxy halo must be assumed to get a probability that is not
extremely small: $P(<\Lambda_{\rm obs})$ = 3.4\% for $M_{\rm DM, th} =
5 \times 10^{11} M_\odot$, though there exist much smaller galaxies.
If we take a smaller threshold of $5 \times 10^8 M_\odot$, the
probability reduces to 0.58\%. Our result using the galaxy formation
model can be understood by the supernova feedback taken into account
in the model; a significant fraction of dark matter mass is
distributed in small mass haloes, but star formation in such haloes is
suppressed by the feedback.

We also tested the possibility that the number of observers depends on
metallicity of galaxies. Introducing a metallicity threshold does not
change the probability $P(<\Lambda_{\rm obs})$ significantly; it
increases from 7.3 to 9.0\% for $Z_{\rm th} = $ 0.1 to 1.0 $Z_\odot$.
This is because low metallicity galaxies are generally small galaxies
(the mass-metallicity relation), and such galaxies do not include a
significant fraction of stellar mass in the universe due to the
supernova feedback. If we assume that the number of observers is
proportional to $N_{\rm life} \propto M_*Z$ of galaxies, we found
$P(<\Lambda_{\rm obs})$ = 9.7\%.

We conclude that a reasonable estimate of the probability to
find a small $\Lambda$ as observed is 7--10\%, which is not
extremely small, based on a realistic model of galaxy
formation. Therefore the anthropic argument is a viable explanation
for the cosmological constant problem.  If, in future, a convincing
theory is established by fundamental physics predicting that only
$\Lambda$ is variable with a flat prior distribution when a universe
is created, the anthropic argument may become the leading candidate
for the solution of the cosmological constant problem.

\section*{Acknowledgements}

TT was supported by JSPS KAKENHI Grant Numbers 15K05018 and 40197778. 
RM was supported in part by MEXT KAKENHI Grant Number 15H05896.





\bsp	
\label{lastpage}
\end{document}